# On estimating superconducting shielding volume fraction from susceptibility in pressurized Ruddlesden-Popper nickelates: Response to arXiv:2602.19282

Yinghao Zhu[1*], Di Peng[2*], Enkang Zhang[1], Bingying Pan[3], Xu Chen[4], Zhenfang Xing[2], Cuiying Pei[5], Feiyu Li[6], Yanpeng Qi[5*], Junjie Zhang[6*], Qiaoshi Zeng[2*], Jian-gang Guo[4*] & Jun Zhao[1*]

[1]*State Key Laboratory of Surface Physics and Department of Physics, Fudan University, Shanghai 200433, China*

[2]*Center for High Pressure Science and Technology Advanced Research, Shanghai 201203, China*

[3]*College of Physics and Optoelectronic Engineering, Ocean University of China, Qingdao, Shandong 266100, China*

[4]*Beijing National Laboratory for Condensed Matter Physics, Institute of Physics, Chinese Academy of Sciences, Beijing 100190, China*

[5] *State Key Laboratory of Quantum Functional Materials, School of Physical Science and Technology, ShanghaiTech University, Shanghai 201210, China*

[6]*State Key Laboratory of Crystal Materials, Institute of Crystal Materials, Shandong University, Jinan, China*

**Abstract**

In a recent preprint (arXiv:2602.19282) [1], the authors questioned the procedure we used to evaluate the demagnetization-corrected superconducting shielding volume fraction in pressurized Ruddlesden–Popper nickelates [2–5]. They further claimed that this methodology has neither been derived nor used previously, and they proposed an alternative normalization scheme. Here we clarify that our evaluation follows directly from the standard magnetostatic self-consistency relation for finite samples and has been widely adopted in the superconductivity literature for decades. We also demonstrate that the discrepancies claimed in Ref. [1] stem from a fundamental flaw in their approach, namely, the assumption that the measured diamagnetic moment is linearly proportional to the superconducting shielding volume fraction in the presence of a finite demagnetization factor $N$. This assumption is not valid for strongly demagnetized, thin disk-like specimens, where the internal field and the measured moment are coupled self-consistently through the demagnetizing field.

To clarify our approach and the confusion raised in Ref. [1], we begin by reviewing the standard magnetostatic self-consistency framework. For a finite specimen of a given geometry, the demagnetizing field modifies the relation between the applied field $H_0$ and the internal field $H$.

$$H = H_0 - NM \qquad (1)$$

where $N$ is the (geometry-dependent) demagnetization factor and $M$ is the volume magnetization. Defining the intrinsic susceptibility $\chi = M/H$ and the measured susceptibility $\chi_0 = M/H_0$, Eq. (1) directly yields the well-known self-consistency relation

$$1/\chi = 1/\chi_0 - N, \qquad (2)$$

In the low-field Meissner regime, after subtracting the normal-state background, the effective superconducting shielding volume fraction can be estimated as $f \approx -\chi$ (in SI units). This is precisely the method we employed in Ref. [2] to obtain the superconducting shielding volume fraction (see Eq. (5) in Ref. [2]), which has been routinely derived and employed in the superconductivity literature (see e.g., Refs. 6–10) for many years.

Eq. (2) can be equivalently rewritten as

$$f = -\chi = \frac{-\chi_0}{1 - N\chi_0} \qquad (3)$$

or

$$\chi_0 = \frac{\chi}{1 + N\chi} \qquad (4)$$

Here, $\chi = -1$ corresponds to 100% superconducting shielding volume fraction.

For the single crystal sample S6 in Ref. 2, the measured DC magnetic moment at 50 GPa and 5 K (in SI units) is $m_{meas} = -1.69 \times 10^{-6}$ emu $= -1.69 \times 10^{-9}$ A m$^2$, and the applied magnetic field is $H_0 = 20$ Oe $= 1591.5$ A m$^{-1}$. At ambient pressure, the sample can be approximated as a cylindrical disk with diameter $A \approx 160$ μm and thickness $C \approx 22$ μm, giving a sample volume $V_0 \approx 4.42 \times 10^{-13}$ m$^3$. Using the relation $N^{-1} \approx 1 + 1.6(C/A)$ [6], we evaluate the demagnetization factor as $N \approx 0.82$. This factor stays essentially constant under hydrostatic pressure, because the lattice contracts proportionally. By applying the cell volume of La$_4$Ni$_3$O$_{10}$ with $Z = 2$ at 0 GPa and 50 GPa, $V_{cell,0\,GPa} = 413.49$ Å$^3$ and $V_{cell,50\,GPa} = 342.84$ Å$^3$ respectively, the sample volume at 50 GPa can be estimated as

$$V_{cylinder,50GPa} = V_{cylinder,0GPa} \times V_{cell,50GPa}/V_{cell,0GPa} \approx 3.66 \times 10^{-13} \text{ m}^3 \qquad (5)$$

The measured susceptibility $\chi_0$ is then obtained directly from the definition

$$\chi_0 = \frac{m_{meas}}{V_{cylinder,50GPa} \times H_0} \approx -2.9 \qquad (6)$$

Substituting the above value into any of Eqs. (2)–(4) yields the intrinsic susceptibility $\chi$. For $N \approx 0.82$, Eq. (3) gives

$$f = -\chi = \frac{-(-2.9)}{1 + 0.82 \times 2.9} \approx 0.86 \qquad (7)$$

Therefore, the superconducting shielding volume fraction at 50 GPa is around 86%. Similarly, applying the same procedure yields a superconducting shielding volume fraction of around 82% for sample S6 at 5 K, 40 GPa and 20 Oe.

The above procedure is based on well-established principles of magnetostatics and longstanding practices in superconductivity research. Numerous publications adopt the same methodology to estimate superconducting shielding volume fractions (see e.g., Refs. [6–35]). It is unfortunate that Ref. [1] overlooked this standard approach and mistakenly claimed that it had never been derived or used before.

**Clarification of Ref. [1]'s method**

Ref. [1] also misrepresents our analysis by ascribing to us a definition that we never used, namely defining the superconducting shielding volume fraction $f'$ as the ratio of the measured magnetic moment to a "Meissner moment". This moment-ratio quantity is **not** the demagnetization-corrected superconducting shielding volume fraction obtained from Eqs. (2)–(4), and it is precisely this misidentification that leads to their underestimation.

As shown in Eq. (2) of Ref. [1], they calculate the fraction $f'$ as the direct ratio between the measured magnetic moment $m_{meas}$ and the calculated Meissner moment $m_{Meissner}$ of an equal-volume, fully shielding specimen, as follows:

$$f' = \frac{|m_{meas}|}{|m_{Meissner}|} \tag{8}$$

Using this normalization, they obtain an apparent "superconducting volume fraction" of around 60% for sample S6 at 50 GPa. The central flaw is that this normalization implicitly assumes linear scaling, $f' \propto m_{meas}$, which is generally not valid when the demagnetization factor $N$ is large (as in thin disk-like crystals). The reason is straightforward: the internal field is not fixed externally, but is determined self-consistently by magnetostatics. In particular, the demagnetizing field satisfies $H_d = NM$, so reducing the shielding fraction reduces the magnitude of $M$, which in turn reduces $H_d$, thereby changing the internal field $H = H_0 - H_d$ that drives the response. This feedback makes $m_{meas}$ a nonlinear function of the effective shielding fraction whenever $N \neq 0$, and therefore a simple ratio $\frac{|m_{meas}|}{|m_{Meissner}|}$ systematically misestimates the superconducting shielding volume fraction.

The difference between the approach of Ref. [1] and the standard demagnetization self-consistency [Eq. (2−4)] can be seen more clearly when expressed in terms of susceptibility rather than magnetic moment.

$$f' = \frac{|m_{meas}|}{|m_{Meissner}|} = -\chi_0(1-N) \tag{9}$$

In addition, Ref. [1] claims that we employed Eq. (3) of Ref. [1] [reproduced below as Eq. (10)] to calculate the superconducting shielding volume fraction without proper attribution:

$$f = \frac{\frac{|m_{meas}|}{|m_{Meissner}|}}{1 + N \times (\frac{|m_{meas}|}{|m_{Meissner}|} - 1)} \tag{10}$$

This accusation is misplaced. In Ref. 2, we stated explicitly that the superconducting shielding volume fraction was obtained from Eq. 2 (corresponding to Eq. 5 of Ref. 2). We did not present Eq. (10) as a new or independent formula.

In our previous email correspondence with the authors of Ref. [1], they preferred to express everything in terms of the measured magnetic moment and the full-Meissner moment. To communicate in their notation, we simply rewrote Eq. 2 in an equivalent "moment form" and obtained Eq. (10). Substituting $\frac{|m_{meas}|}{|m_{Meissner}|} = -\chi_0(1-N)$ back into Eq. (10), one immediately recovers Eqs. 2–4, exactly.

**Scope and applicability of the single-demagnetization-factor (single-$N$) analysis**

The estimate of superconducting shielding volume fraction is based on the standard magnetostatic self-consistency for a specimen whose macroscopic response can be described by a single demagnetization factor $N$. In this framework, the measured low-field response is described by a volume-averaged (effective) susceptibility, and it does not require microscopic uniformity on short length scales. Experimentally, our single crystals exhibit high structural and chemical quality, as supported by scanning transmission electron microscopy, energy-dispersive spectroscopy, and single-crystal neutron and X-ray diffraction. Under hydrostatic pressure, we further observe a robust zero-resistance state and a mutually consistent pressure evolution of the resistive and diamagnetic superconducting transitions. Together, these observations support the applicability of a single-$N$ framework for describing the specimen's low-field macroscopic response. This approach follows well-established magnetostatics and long-standing superconductivity practice, and has been routinely used in the superconductivity literature for decades [6-35].

With this in mind, the "50% superconducting volume" test cases in (Samples A and B in Ref. [1]) are not within the applicability of the single-$N$ self-consistent relation. Specifically, Samples A/B are explicitly inhomogeneous magnetostatic boundary-value problems: for phase-separated core–shell/lamella constructions, the internal field $H$ is strongly position dependent, and the composite specimen cannot, in general, be characterized by a single demagnetization factor $N$ in the sense assumed by the self-consistent relation. Such macroscopic phase-separated morphologies are also inconsistent with our experimental characterization.

Moreover, in calculations in Ref. [1], the authors treat the superconducting sub-region as if it were an isolated disk in free space with its own demagnetization factor $N_{sub}$, and they compute its moment using the formula for a standalone fully shielded body, then normalize by the Meissner moment of the full disk and insert the full-disk $N$ back into Eq. (10). However, a superconducting lamella or core embedded in a normal host does not experience the same magnetostatic environment as a free isolated disk. Its moment cannot be obtained by simply applying the isolated-body Meissner formula with $N_{sub}$, the demagnetizing field is determined by the full composite geometry and continuity conditions at the SC/normal boundary. This suggests a "geometric 50%" construction cannot be used as a consistency check for a formula derived in the single-$N$ frame work.

Accordingly, the toy constructions presented in Ref. [1] under their chosen procedure do not demonstrate that the demagnetization self-consistency is incorrect. Instead, they simply reflect a well-established fact: the single-$N$ response framework is not intended to recover arbitrary, morphology-dependent geometric phase fractions in intentionally phase-separated specimens.

**Unit conventions and notation**

Regarding the unit system, in Ref. [2] we used Gaussian (cgs) units for the susceptibility plot primarily for experimental convention, since SQUID magnetometry is commonly reported in emu and Oe. Importantly, the choice of unit system does not affect any conclusions, because all quantities can be converted to SI unambiguously. In the present article, we have converted everything to SI units, and we consistently obtain the same superconducting shielding volume fraction for sample S6 at 50 GPa, namely $f = -\chi = 0.86$ (86%). Expressed in cgs form, this corresponds to $f = -4\pi\chi = 0.86$ (86%), i.e., the same superconducting shielding volume fraction written using the conventional $4\pi\chi$ notation.

As for the use of the symbol **M** in Ref. 2, SQUID magnetometry directly measures the magnetic moment (reported in emu by the instrument software). In Fig. 3 of Ref. [2], **M** (emu) was used as a shorthand for the measured moment plotted in emu. Because magnetization would instead carry units of emu cm$^{-3}$ (or A m$^{-1}$), the quantity plotted in Fig. 3 unambiguously corresponds to the magnetic moment.

**Conclusion**

In summary, the estimation of the superconducting shielding volume fraction in pressurized Ruddlesden-Popper nickelates is conducted within the well-established magnetostatic self-consistency framework, as outlined in Eqs. (2)–(4). This framework ensures that the demagnetization-corrected superconducting shielding volume fractions are consistent with long-standing experimental practices in the superconductivity community. The discrepancies claimed in Ref. [1] arise from a fundamental flaw in their methodology, particularly the assumption that the measured diamagnetic moment is linearly proportional to the superconducting shielding volume fraction in the presence of a finite demagnetization factor $N$. This assumption does not hold for strongly demagnetized disk-shaped specimens, where the internal field and the resulting magnetization are not linearly proportional to the shielding fraction. Therefore, our approach, grounded in magnetostatic self-consistency, remains the correct and widely accepted method for estimating superconducting shielding volume fractions.

*Email: Y.Z. (yinghaozhu@fudan.edu.cn), D.P. (pengdi@sharps.ac.cn), Y.Q. (qiyp@shanghaitech.edu.cn), J.J.Z. (junjie@sdu.edu.cn), Q.Z. (zengqs@hpstar.ac.cn), J.G. (jgguo@iphy.ac.cn), J.Z. (zhaoj@fudan.edu.cn).